\documentstyle[11pt,emulateapj]{article}
\begin{document}

\title{Constraints on primordial black holes and primeval density 
  perturbations from the epoch of reionization}

\author{Ping He and Li-Zhi Fang}

\affil{Department of Physics, University of Arizona, 
         Tucson, AZ 85721, USA}

\begin{abstract}

We investigate the constraint on the abundance of primordial black holes 
(PBHs) and the spectral index $n$ of primeval density perturbations given by 
the ionizing photon background at the epoch of reionization. Within the 
standard inflationary cosmogony, we show that the spectral index 
$n$ of the power-law power spectrum of primeval density perturbations 
should be $n<$1.27. Since the universe is still optical thick 
at the reionization redshift $z\sim 6$ - 8, this constraint is independent 
of the unknown parameter of reheating temperature of the inflation.
The ionizing photon background from the PBHs can be well approximated by a
power law spectrum $J(\nu)\propto{\nu}^3$, which is greatly different 
from those given by models of massive stars and quasars.

\end{abstract}


\section{Introduction}

Primordial black holes (PBHs) form from the gravitational collapse when 
an initially dense region on super-horizon scales enters the horizon 
during the early radiation-dominated era (Carr \& Hawking 1974, Carr 1975).
In the inflationary universe, the abundance of PBHs is extremely
sensitive to the spectral index $n$, which specifies the dependence
of the power spectrum of primordial density fluctuations on the
comoving wavenumber $k$, $|\delta_k|^2\propto k^n$. Thus, a constraint
on the abundance of PBHs will also yield a constraint on the index $n$. 
 
The simplest constraint on the abundance of PBHs is given by the condition
$\Omega_{bh} \leq 1$, i.e., the mass density of the PBHs should be 
less than the critical density of the universe. This condition results in
$n <1.5$ regardless of whether a black hole can evaporate completely, or 
ends with a Planck relic (Kim \& Lee 1996). More effective constraints
can be derived by comparing the particle emission from PBH evaporation with
observed background. For instance, the cosmic background of $\gamma$-ray 
spectrum on $\sim 100$ Mev places a tightest constraint on the PBH density
parameter $\Omega_{bh} \leq 10^{-8}$ (MacGibbon \& Carr 1991), which yields 
the index $n<1.25$ (Kim, Lee \& MacGibbon 1999). However, these constraints
depend on an additional parameter, the reheating temperature $T_{rh}$, 
or equivalently, the minimum mass of PBHs $M_{min}$, which is not well 
determined. 

Recently, the epoch of the onset of the cosmic reionization is explored with
absorption spectrum of high redshift ($5.7<z<6.3$) SDSS quasars (Fan et al. 
2001, and references therein). The existence of the complete Ly$\alpha$ 
Gunn-Peterson trough in the absorption spectra of high redshift quasars 
indicates that the dark age probably ended at the redshift $z\simeq 6 - 8$. 
The ionizing photon background in the IGM at $z\sim 6$ is found to be more 
than 20 times lower than that at $z\sim3$. These results motivate us to 
explore the constraints on the abundance of PBHs, and then on the primordial 
density perturbations with the low energy photon background at the epoch
of reionization. We will show that this constraint is basically independent of
the uncertain parameters $T_{rh}$ or $M_{min}$. PBHs have been considered
as a possible source of the cosmic reionization (Gibilisco 1996, 1997), but in 
these works, the mass function of PBHs is treated as free parameters. It cannot 
be used to constrain $n$.     

This {\it Letter} is organized as follows. In \S 2, we briefly present the
methods of calculating the mass function of the PBHs and the ionizing photon
background from PBH's evaporation. The numerical results of the constraints 
on $n$ from the observed high redshift ionizing background are given in \S 3. 
\S 4 are discussions and conclusions.

\section{Methods}

\subsection{Mass function of PBHs}

For Gaussian primeval density perturbations, the mass function 
$n_{bh}(M_{bh})$ of PBHs at redshift $z$ can be calculated by the same 
way as calculating the mass function of collapsed massive halos with 
the Press-Schechter formalism (Press \& Schechter 1974). Accordingly, 
the number density of regions with density contrast from $\delta$ to $\delta 
+d\delta$ and mass $M$ to $M+dM$ at the time $t_i$ of the onset of 
radiation-dominated epoch is 
\begin{eqnarray}
\lefteqn{n(M, \delta ) dMd\delta  = } \\ \nonumber
 & & \hspace{-3mm} 
  \sqrt{\frac{2}{\pi}}\frac{\rho_i}{M^2} 
       	\frac{1}{\sigma_R} 
\left | \frac{\partial \ln \sigma_R}{\partial \ln M}\right |
      \left|\frac{\delta^2}{\sigma_R^2}-1\right |        
     e^{-\frac{\delta^2}{2\sigma_R^2}} dM d\delta, 
\end{eqnarray}
where $\rho_i$ is the cosmic mean mass density at $t_i$, $M$ the 
mass within $R$, 
and $\sigma_R$ the variance of the probability distribution 
function of initial Gaussian mass field smoothed by scale $R$.  
Since the mass function is normalized at $\rho_i$, i.e. 
$\int\int Mn(M, \delta)dMd\delta= \rho_i$, eq.(1) gives the 
physical number density of region $(M, \delta)$ at $t_i$. 
The comoving number density is 
$n(M, \delta )_{com}=n(M, \delta )(a_i/a_0)^3$, where
$a(t)$ is the cosmic factor.

It has been shown (Carr 1975) that when a region $(M,\delta)$ 
enters the horizon at $t_h$, i.e. when $M$ is about the
same as horizon mass at $t_h$, this region will collapse to a 
black hole if the initial density contrast $\delta$ satisfies
\begin{equation}
\gamma(M/M_i)^{-2/3} \leq \delta \leq (M/M_i)^{-2/3},
\end{equation}
where parameter $\gamma \simeq 1/3$ and $M_i$ is the horizon mass 
at $t_i$, i.e. the horizon mass at the beginning of the 
radiation-dominated. The mass of a black hole formed from this 
region is equal to $M_{bh}=\gamma^{1/2}M_i^{1/3}M^{2/3}$. Thus, 
the mass function of PBHs is
\begin{eqnarray}
\lefteqn{ n_{bh}(M_{bh}) =
  \frac{dM}{dM_{bh}}
  \int_{\gamma(M/M_i)^{-2/3}}^{(M/M_i)^{-2/3}} 
          n(M, \delta )d\delta =  } \\ \nonumber 
 & & \hspace{-8mm} \frac{3}{2}
  \sqrt{\frac{2}{\pi}}
  \frac{\gamma^{3/4}\rho_iM_i^{1/2}}{m_{bh}^{5/2}} 
  \int_{\gamma}^{1} 
  \left |\frac{\partial \ln \sigma_R}{\partial \ln M}\right |
  \left|\frac{\delta_h^2}{\sigma_h^2}-1\right |   
  e^{- \frac{\delta_h^2}{2\sigma_h^2}} \frac{1}{\sigma_h}d\delta_h,
\end{eqnarray}
where $\delta_h=\delta (M/M_i)^{2/3}$ is the density contrast when the 
region $(M, \delta)$ crosses the horizon at $t_h$, and variance
$\sigma_h=\sigma_R (M/M_i)^{2/3}$. Since $M \geq M_i$, the minimum 
mass of the PBH is $\gamma^{1/2}M_i$. 

For a power law initial density perturbation with spectral index $n$, 
i.e. $\sigma_h \propto M^{(1-n)/6}$, we have 
$\partial \ln \sigma_R/\partial \ln M =(n+3)/6$, and eq.(3) becomes 
(Kim \& Lee 1996)
\begin{eqnarray}
\lefteqn{n_{bh}(M_{bh})= } \\ \nonumber
 & & \frac{n+3}{4}\left(\frac{2}{\pi}\right)^{1/2}\gamma^{7/4}\rho_i
  M_i^{1/2}M_{bh}^{-5/2} \sigma_h^{-1} e^{-\gamma^2/2\sigma_h^2}.
\end{eqnarray} 
As usual, the variance $\sigma_h$ will be normalized with the COBE 
observation (e.g., Fang \& Xu 1999, Bugaev \& Konishchev 2001). 

In eq.(4), only the factors $\rho_i=3/32\pi G t_i^2$ and 
$M_i=(4\pi/3)(ct_i)^3\rho_i$ are dependent on $t_i$. Thus,
$n_{bh}(M_{bh}) \propto t_i^{-3/2}$. On the other hand, 
$(a_i/a_0)^3 \simeq (1+z_{eq})^3(t_i/t_{eq})^{3/2}$, where $t_{eq}$ 
and $z_{eq}$ are, respectively, the time and the redshift at 
the end of the radiation-dominated era. Thus, the comoving mass 
function $(a_i/a_0)^3n_{bh}(M_{bh})$ is independent of $t_i$. 
  
\subsection{Ionizing photon background from PBH's evaporation}

For a Schwarzschild black hole $m$, the Hawking emission rate 
for particles with spin $s$, energy from $E$ to $E+dE$ and per
degree of particle freedom is  
\begin{equation}
 \frac{dN}{dt}=  
   \frac{1}{2\pi \hbar}\frac{\Gamma_s dE}{e^{E/k_BT(m)}-(-1)^{2s}},
\end{equation}
where $T(m)=\hbar c^3/8\pi G k_B m$ is the temperature of 
black hole $m$, and the dimensionless parameter $\Gamma_s$ is the 
absorption probability of spin $s$ particle by a black hole $m$ 
(MacGibbon \& Webber 1990; Page 1975). For low energy photon 
emission, i.e., $h \nu \ll k_bT(m)$, 
$\Gamma=128G^4m^4\nu^4/3c^{12}$, and eq.(5) gives 
\begin{equation}
\frac{d^2N}{d\nu dt}=f(\nu, m)=
  \frac{128G^4m^4\nu^4}{3c^{12}[e^{h\nu/k_BT(m)}-1]}.
\end{equation}

The evaporation leads to the decreasing of black hole mass with 
the rate
\begin{equation} 
\frac{dm}{dt} = -\frac{\alpha(m)}{m^2}
\end{equation}
where $\alpha(m)$ takes into account all degrees of freedom of 
evaporated particles (MacGibbon 1991). Since $\alpha(m)$ is not 
a strong function of $m$, the time dependence of $m$ for a black 
hole with initial mass $M_{bh}$ at $t_h$ is approximated by
\begin{equation}
m(M_{bh}, t) \simeq [M^3_{bh} -3\alpha(m) t]^{1/3},
\end{equation} if $t \gg t_h$.    
 
Thus, the background photon energy flux $J(\nu, z_{obs})$ per unit 
frequency interval and per solid angle at redshift $z_{obs}$ 
is as follows:
\begin{eqnarray}
\lefteqn{\frac{J(\nu, z_{obs})}{h\nu}= 
  \frac{c}{4\pi}\int^{t(z_{obs})}_{t_{eq}} 
   dt \left (\frac{a_{obs}}{a}\right )
   \left(\frac{a_i}{a_{obs}}\right )^3 } \\ \nonumber
 & & \hspace{-5mm}  \int_{M_{min}}^{M_{max}} dM_{bh} n_{bh}(M_{bh})
   \exp[-\tau_{eff}(\nu, z_{obs}, z) ]
  \\ \nonumber
 & &   f[\nu (1+z)/(1+z_{obs}), m(M_{bh},t)],
\end{eqnarray}
$M_{min}$ is given by $\max[\gamma^{1/2}M_i, m(t)]$, where $m(t)$ is the
solution of $m(t)=\{ 3\alpha[m(t)]t\}^{1/3}$. The integral upper 
limit $M_{max}$ in eq.(9) in not important, as the mass function 
$n_{bh}(M_{bh})$ is rapidly decreasing with the increasing of $M_{bh}$. 
$\tau_{eff}(\nu, z_{obs}, z)$ is the opacity for 
photons emitted at $z$ and observed at $z_{obs}$ at the frequency 
$\nu$.
 
\section{Results}

Using eq.(9), we calculate the ionizing photon background at 
redshift $z_{obs} \sim 6$ - 8. The opacity $\tau_{eff}$
for photons observed at the hydrogen Lyman edge ($h\nu_0$ =13.6 ev) 
is dominated by HI photoelectric absorption. We will 
use the $\tau_{eff}$ estimated by the so-called high absorption 
model (Meiksin \& Madau, 1993). It is 
\begin{eqnarray}
\lefteqn{ \tau(\nu_0, z_{obs}, z) \simeq } \\ \nonumber
 & & 0.0097(1+z_{obs})^{1.56}[(1+z)^{1.84}-(1+z_{obs})^{1.84} ]
   \\ \nonumber
 & & - 0.0068(1+z_{obs})^{1.56}[(1+z)^{0.4}-(1+z_{obs})^{0.4} ]
   \\ \nonumber
  & & 8.06\times 10^{-5}[(1+z)^{3.4}-(1+z_{obs})^{3.4} ].
\end{eqnarray}
The optical depth given by eq.(10) is very short. For $z_{obs}=6$, 
the mean transmission $e^{-\tau_{eff}}$ is less than 
$10^{-4}$ at $z\simeq 6.5$. That is, the effective flux $J(\nu_0)$
observed at $z=6$ is only contributed by PBHs localized within the
spatial range from $z=6$ to 6.5. For $z=8$, this range is about 
$z=8$ to 8.25. Actually, eq.(10) is found from a fitting with 
observations at $z \sim 3$ - 4. However, there is no direct 
observation of the hydrogen Lyman edge absorption at redshifts 
$z>6$. Nevertheless, eq.(10) is consistent with the Ly$\alpha$ 
transmission $e^{-\tau} = 0.004\pm 0.003$ in the redshift range 
$[5.95-6.15]$ (Becker et al. 2001). Hence, it would be reasonable to 
estimate the optical depth at $z=6$ - 6.5 by eq.(10).

The latest observation of the complete Gunn-Peterson troughs in 
the absorption spectrum of a $z$=6.28 quasar indicates that the 
photoionization rate at $z \sim$6 is lower than 
8.0$\times10^{-14}$s$^{-1}$ for Ly$\alpha$, or 
2.0$\times10^{-14}$s$^{-1}$ for Ly$\gamma$ (Fan et al. 2001). 
Assuming the ionizing photons are from massive stars 
$J(\nu) \propto \nu^{-5}$ (Barkana \& Loeb 2001),
one can then find that the upper limits to the ionizing photon 
background $J_{-21}\equiv J(\nu)/10^{-21}$ 
($J(\nu)$ in unit of erg cm$^{-2}$ sec$^{-1}$Hz$^{-1}$sr$^{-1}$)
at photon energy 13.6ev are, respectively, $5.3\times10^{-2}$ and
$1.3\times10^{-2}$ for Ly$\alpha$, and Ly$\gamma$, and at 24ev,
are $3.1\times10^{-3}$, $7.8\times10^{-4}$ for Ly$\alpha$, and Ly$\gamma$,
respectively. 

Figure 1 presents the PBH contributed ionizing photon background 
 $J_{-21}$ at $z=6$ as a function of the index $n$ of the 
primordial fluctuation power spectrum. The upper limits to
the ionizing photon background yield $n < 1.273$ for $E$=13.6ev, 
and $n < 1.270$ for $E$=24 ev. 

It should be pointed out that these constraint basically are 
independent of $t_i$. Eq.(9) shows that $J(\nu_0)$ depends on $t_i$ 
only via $M_{min}$, the truncation of PBH mass function at small 
mass side. On the other hand, the optical depth at $z=6$ is very 
small. Only PBHs, which can survive to $z\simeq 6$, has contribution
to $J(\nu_0)$. Thus, $J(\nu_0)$ is independent of $M_{min}$, if
$M_{min} \leq 10^{12}$ g, as those PBHs have already evaporated at 
$z\simeq 6$ [eqs.(7) and (8)]. In inflationary universe, 
$t_i=0.301 g_*^{-1/2}M_{pl}/T^2_{rh}$ (Kolb \& Turner 1990). Thus,
the $M_{min}$-independence leads to also $T_{rh}$-independence. 

The $M_{min}$ or $T_{rh}$-independence of $J(\nu_0)$ can be also 
seen from Fig. 2, which plots $dJ(\nu_0, z_{obs})/dm$ vs. $m$. 
It shows that $J(\nu_0)$ at $z=6$ is mainly contributed by the 
evaporation of PBHs with mass $m\simeq 2\times 10^{14}$ g. Therefore,
$J(\nu_0)$ is independent of $M_{min}$ if it is less than 
$2\times 10^{14}$ g. The peak $2\times 10^{14}$ g. is also 
consistent with the assumption of $h \nu \ll k_bT(m)$ used in 
eq.(6). 

We calculate the flux $J(\nu, z)$ at $z=6$ in the frequency range 
from $\nu$ = 3$\times 10^{15}$ to $6 \times 10^{15}$, which
corresponds to photon energy range from 13.6 to 24.6 ev (HeI ionization). 
In this frequency interval one can still approximately use the optical 
depth $\tau_{eff}$ of eq.(10). The result is shown in Fig.3. The 
flux $J(\nu)$ can be well approximated by a power law spectrum 
$J(\nu)\propto\nu^{3}$. This spectrum is very different from the 
models of massive stars ($J(\nu)\propto\nu^{-5}$) and quasars
($J(\nu)\propto\nu^{-1.8}$, see Madau, Haardt, \& Rees 1999),
for both of which, the flux decreases with the increasing of $\nu$.
This is because $J(\nu)$ 
are mainly from the low energy tail of the photon emission by 
10$^{14}$g black holes, whose temperature $T(m) \gg h\nu_0/k_b$. 
In the low energy tail, the intensity is increasing with $\nu$. 

In all the above calculations, the photons produced from the 
fragmentation of the PBH-emitted quarks and gluons are not considered
(MacGibbon 1991). Since $J(\nu_0)$ is given by local PBHs, we need only
to consider the fragmentation of quarks and gluons into low energy 
($\sim 10$ - 20 ev) photons. This problem is still quite unclear and
uncertain. Since the fragmentation effect will lead to an 
increase of $J(\nu_0)$. Therefore, all above-mentioned upper limits 
on $n$ will be held. 

\section{Discussions and Conclusions}

In the standard inflationary model with a power-law power spectrum of
the primeval density perturbations, we calculated the ionizing background 
flux of the PBHs. With the latest observed ionizing photon background 
flux at the redshift of reionization, we find that the spectral index of the 
primeval density perturbations should be $n<1.27$, assuming that the background
photon flux are emitted by massive stars. The difference is negligible if
the background flux coming from quasars. If the parameter 
$\gamma$ is taken to be 0.6 (Niemeyer \& Jedamzik 1998), we have $n<$1.29.
We also calculate $J(\nu)$ at $z=8$. The result is almost the same as
Fig. 3. However, there is no available data on the $J(\nu_0)$ at this
redshift. If the reionization is really onset at $z \simeq 6$ - 8, 
$J(\nu_0)$ at $z=8$ would be significantly less than that at $z =6$,
one may have more tight constraint. All these constraints are independent 
of the reheating temperature $T_{rh}$, if $T_{rh} >10^{10}$ Gev. The abundance
of black hole given by this constraint is $\Omega_{bh}< 10^{-4}$, which 
is much stronger than the critical density constraint $\Omega_{bh}<1$. 

The current observations of the angular power spectrum of the temperature 
fluctuations of cosmic microwave background radiation (CMB) are found 
to be able to fit with a $n=1$ power spectrum on angular scales from 
tens of degrees to sub-degrees. The PBH's constraint $n<1.27$ shows that 
primeval density perturbations might still be scale invariant on scales
much less than the direct observation of the CMB. Moreover, in the CMB
data fitting, the parameters of the spectral index $n$ and the optical 
depth of the CMB photons to the last scattering surface $\tau_c$ are 
degenerated (Stompor et al. 2001). That is, even when the fitted result is 
$n_{fitting} \simeq 1$, the original index may be $n>1$ if $\tau_c >0$. 
On the other hand, the constraint on $n$ given by the PBH is independent 
of the optical depth $\tau_c$.

Thanks to Hongguang Bi for stimulating discussions. PH thanks
M. Gibilisco for sending him her papers and helpful discussions. PH is
supported by a Fellowship of the World Laboratory.

\figcaption{The PBH's ionizing background flux $J_{-21}$ as a function 
  of spectral index $n$ of the primeval density perturbations at $z$=6:
  1. photon energy $E$=13.6 ev (solid); 2. 24 ev (dashed).
  The observed upper limits to $J_{-21}$ at $E$=13.6 and 24 ev are shown,
  respectively, by the shaded areas at top-right and bottom left. 
  The upper and lower boundaries of the two areas refer to the limits 
  given by Ly$\alpha$ and Ly$\gamma$ absorption troughs. These limits 
  are estimated with the model of massive stars.
}

\figcaption[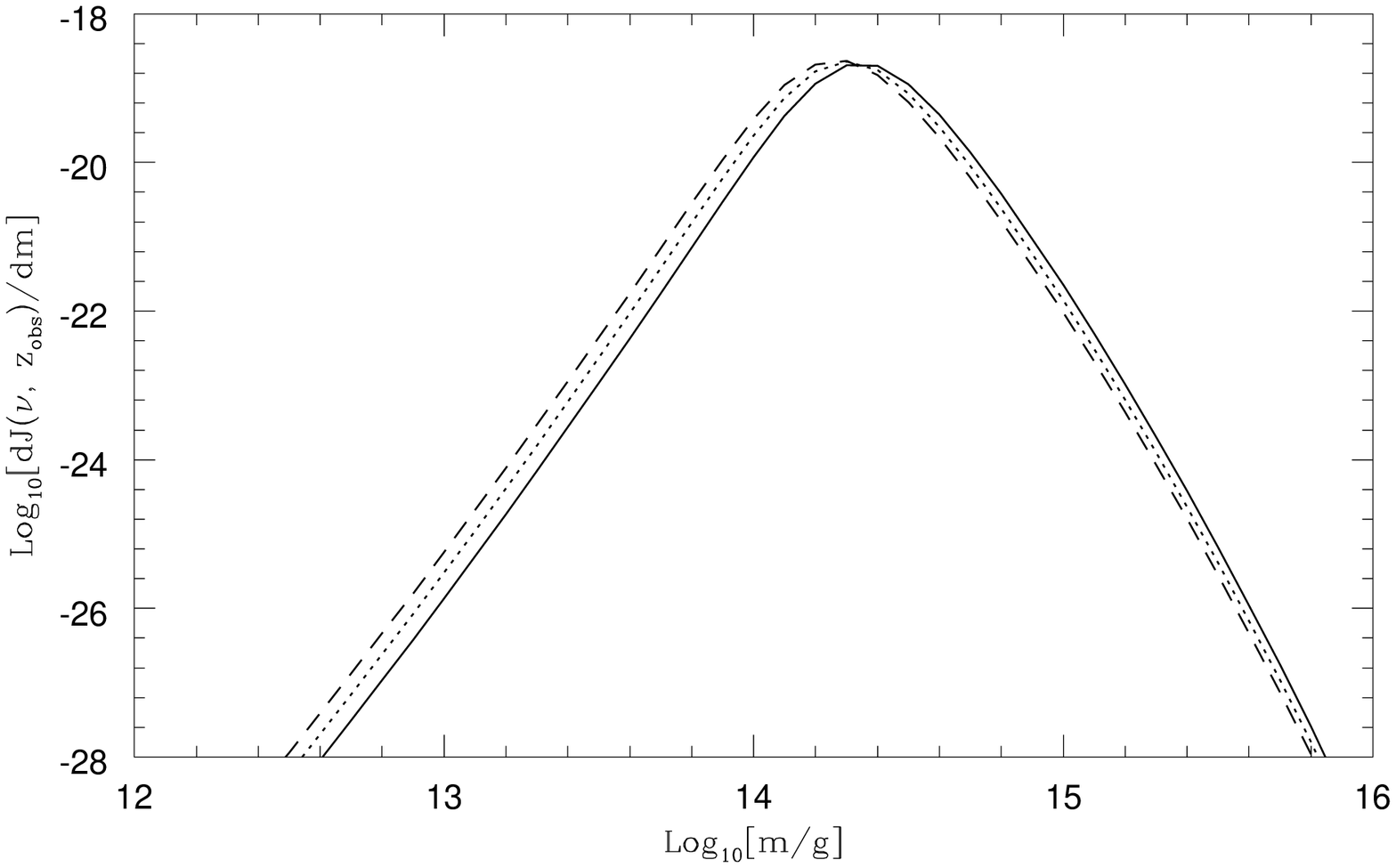] {The differential ionizing background flux 
$dJ(\nu_0, m)/dm$ vs. $m$, with the power
spectrum index $n$=1.27. The solid, dotted, and dashed lines refer to
cases of $z$=6, 8, and 10, respectively.
}

\figcaption[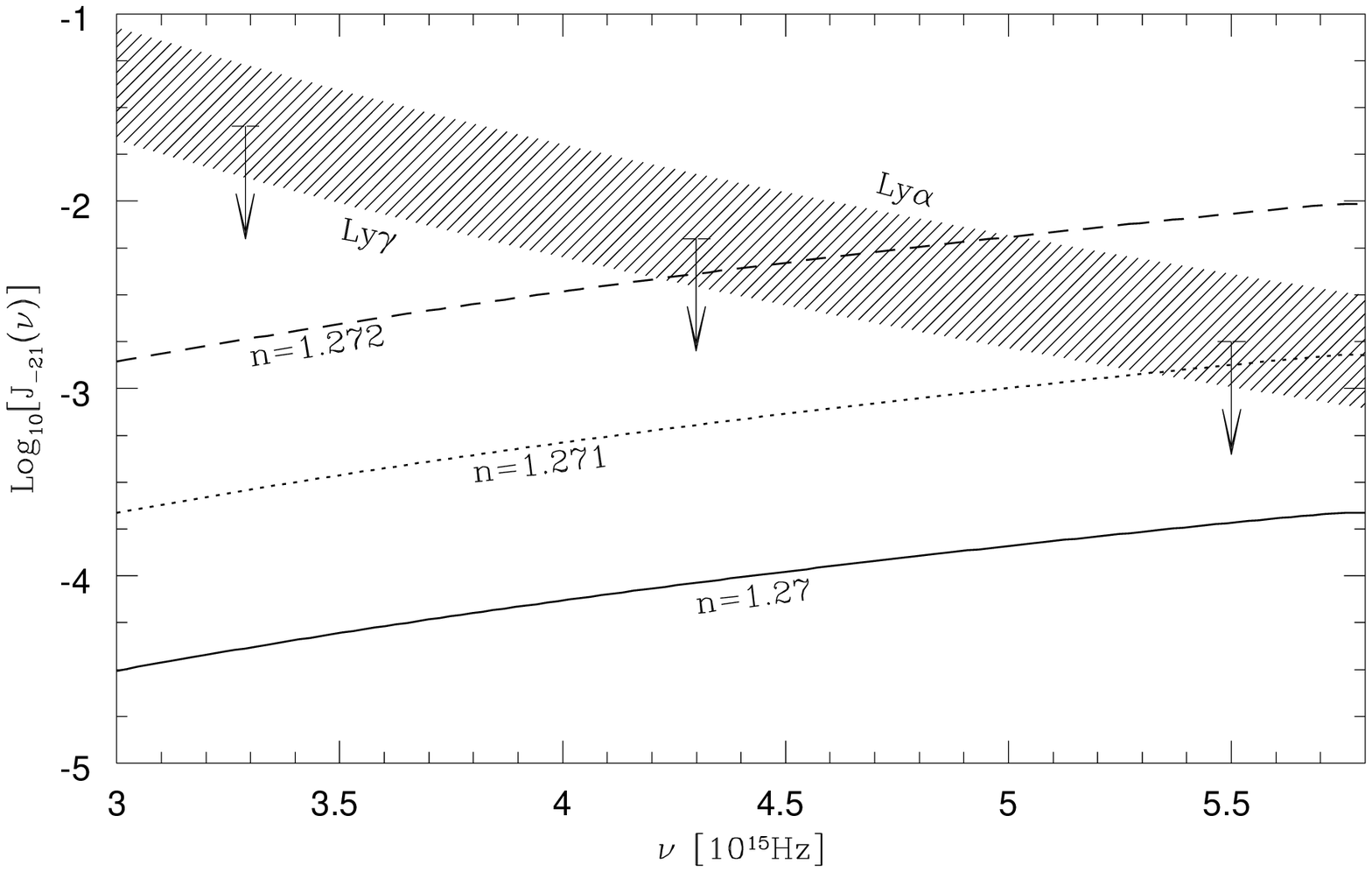]{The spectrum of ionizing photon produced by
 PBHs. The shaded area is upper limits to the ionizing photon spectrum
 given by Ly$\alpha$, and Ly$\gamma$ in the massive star model.
}

\end{document}